\documentclass[fleqn,12pt,twoside]{article}
\usepackage{espcrc1,epsfig}

\usepackage{graphicx}
\usepackage[figuresright]{rotating}

\newcommand{\beq}{\begin{equation}}
\newcommand{\eeq}{\end{equation}}

\newcommand{\AmS}{{\protect\the\textfont2
  A\kern-.1667em\lower.5ex\hbox{M}\kern-.125emS}}

\title{What are we learning from RHIC?\footnote{Invited talk given at the XX International Symposium on 
Lattice Field Theory ``Lattice 2002'', Cambridge, MA, June 24 - 29, 2002.}}

\author{D. Kharzeev\address[BNL]{Nuclear Theory Group, \\
        Physics Department,\\ 
        Brookhaven National Laboratory, \\ 
        Upton, New York 11973-5000, USA}%
        \thanks{Work supported by the U.S. Department of Energy under Contract No. DE-AC02-98CH10886.}}

\begin{document}

% typeset front matter
\maketitle

\begin{abstract}
This talk is an attempt to summarize some of the first results obtained at RHIC. 
I discuss the significance of these measurements for establishing the properties of 
hot and dense QCD matter and for understanding the dynamics of the theory  
at the high parton density, strong color field frontier. 
%The hopes and expectations for the future are discussed as well.     

\end{abstract}

\section{Introduction}
\subsection{What is RHIC?}
Relativistic Heavy Ion Collider (RHIC) at Brookhaven National Laboratory 
on Long Island, New York, began operation in 2000 culminating over ten years 
of development and construction, 
and a much longer period of theoretical speculations about the properties 
of hot QCD matter produced in nuclear collisions in the collider regime.  

 RHIC's $2.4$-mile rings contain superconducting 
magnets, which operate at minus $451.6$ degrees Fahrenheit, $4.5$ degrees above the absolute zero. 
RHIC collides two intersecting 
heavy ion beams at center-of-mass energy of up to 200 GeV/A (at luminosity of up to 
$10^{26} sec^{-1} cm^2$, which can be further increased in the future), and 
polarized proton beams at c.m.s. energy of up to 500 GeV. 
 The total energy 
in the gold--gold collision thus reaches 40 TeV, which is at present 
the World's record collision energy. 
In the $pp$ mode, the unique possibility offered by RHIC for the first time is the study of 
double spin asymmetries and other spin observables. 

\subsection{RHIC experiments}
There are currently five experiments at RHIC: BRAHMS, PHENIX, PHOBOS, STAR, 
which can operate in both heavy ion and $pp$ mode, and PP2PP, which aims exclusively at the study 
of elastic and diffractive $pp$ interactions. These experiments are different in their focus; 
however all of them also provide the information about the basic properties of the collisions. 
The data from different collaborations can thus be cross--correlated and independently verified. 

The BRAHMS experiment is designed to measure charged hadrons over a wide range of rapidity and 
transverse momentum. PHENIX is a large versatile experiment designed to study the 
production of charged and neutral hadrons, leptons, and photons.
The PHOBOS experiment is aimed at measuring the global properties of nuclear collisions, and 
at the study of fluctuations and correlations in the production of hadrons.  
STAR, one of the two large--scale RHIC experiments, focuses on
measurements of the production of various hadron species over a large solid angle and is  
well suited, in particular, for the study of multi--particle correlations and fluctuations 
on the event--by--event basis.

\subsection{RHIC physics: QCD}

RHIC is a machine entirely dedicated to the study of Quantum Chromo--Dynamics -- the theory 
of strong interactions. This includes 
the study of matter at unprecedented energy densities and understanding the structure of the 
proton, in particular, its spin composition. Since the spin program at RHIC at the time of 
this talk is at the very beginning, I will concentrate in this talk entirely on the 
heavy ion program. 

Asymptotic freedom of QCD \cite{Gross:1973id}, \cite{Politzer:fx} 
ensures that the dynamics of quarks and gluons at sufficiently high 
density can be addressed by weak coupling methods. This includes both the thermalized quark--gluon plasma 
at high temperature and the 
wave functions of the colliding nuclei described, at small Bjorken $x$,  
by parton saturation and the Color Glass Condensate \cite{GLR,MUQI,BM,MV}.

\section{Looking at the first RHIC data}

\subsection{Global observables}

Global observables are the most general characteristics of the collision, including particle 
multiplicity, its dependence on the centrality of the collision and on rapidity, and 
azimuthal distribution of the produced particles. The centrality is determined 
by the impact parameter in the collision; since this is not a quantity which can be measured 
directly, centrality is usually determined in terms of a certain cut in the multiplicity 
distribution; e.g., a $0-10 \%$ centrality cut means that out of a given sample, $10 \%$ of the events 
which have the highest multiplicity have been selected. 
It is convenient to characterize centrality in terms of the number of ``participants'' -- 
the nucleons which underwent an inelastic interaction in a given collision. Glauber theory 
\cite{Glauber} can be 
used to correlate a certain centrality cut with an average number of participants (for an explicit set 
of formulae for nuclear collisions, see e.g. \cite{Kharzeev:1996yx}, \cite{Kharzeev:2000ph}).
This procedure can be independently verified by measuring the energy carried forward by spectator 
neutrons; for that purpose all RHIC experiments are equipped by Zero Degree Calorimeters.  
   
\subsubsection{Multiplicity}

Multiplicity in heavy ion collision tells us which 
fraction of the collision energy is inelastically transferred to secondary particles. 
\begin{figure}[htb]
\epsfig{file=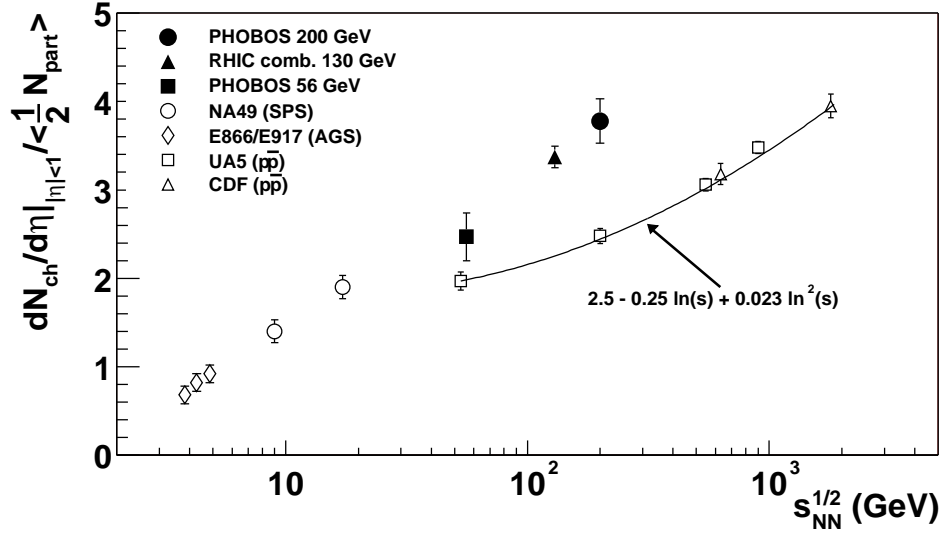,width=13cm}
\caption{Charged multiplicity per participant pair in central Au Au collisions as measured at RHIC 
is compared to  
the multiplicity in $pp$ and $\bar{p}p$ collisions; from \cite{Back:2001ae}.}
\label{fig:phobmult}
\end{figure}
Theoretical expectations for hadron multiplicities at RHIC varied by factor of five, and 
the experimental verdict was thus eagerly awaited. After the commissioning of the machine, the first 
multiplicity results did not take long to come; they are shown on Fig. \ref{fig:phobmult} in 
comparison to the multiplicity previously measured in  $pp$ and $\bar{p}p$ collisions. 
The measured multiplicity appeared much smaller than most theoretical predictions. 
Is this disappointing? To answer this question, let us recall that, by the very definition, 
an incoherent superposition of {\it independent} nucleon--nucleon collisions 
yields multiplicity equal to the number of collisions times the multiplicity in $NN$ collision. 
This trivial statement holds also in the presence of elastic rescatterings. Indeed,      
according to so-called AGK cutting rules \cite{agk} of multiple scattering theory, the
nuclear cross section is given by
\beq
E\frac{d^3\sigma^a_{AB}}{d^3p}=T_{AB}(\vec b)E\frac{d^3\sigma^a_{NN}}{d^3p}, 
\eeq
where the nuclear overlap function is 
\beq
T_{AB}(\vec b)=\int d^2s T_A(\vec s)T_B(\vec b-\vec s),
\eeq
and the nuclear thickness function $T_A(\vec b)=\int_{-\infty}^\infty dz \rho_A(\vec b,z)$ 
is the integral over the
nuclear density.
Integration over impact parameter $b$ yields
\beq
E\frac{d^3\sigma^a_{AB}}{d^3p}=AB\;E\frac{d^3\sigma^a_{NN}}{d^3p},
\eeq
and correspondingly the particle multiplicity would scale as 
\beq\label{eq:nch}
\frac{dn}{d\eta}=AB\;\frac{1}{\sigma^{in}_{AB}}
\frac{d\sigma_{NN}}{d\eta} \ \sim A^{4/3} B^{4/3} \ \frac{dn_{NN}}{d\eta}. 
\eeq
Using the numbers of collisions ($\simeq 1050$) and participants ($\simeq 340$) in central 
($0-6 \%$ centrality cut) Au Au collisions from Glauber 
 model calculations \cite{Kharzeev:2000ph}, we would thus conclude that 
$Au Au$ multiplicity per participant pair should exceed $NN$ multiplicity by factor of $3$. 
Instead, the data at the highest RHIC energy of $\sqrt{s} = 200$ GeV show the difference of 
only about $50 \%$. Given that any inelastic rescatterings in the final state can only 
increase the multiplicity, we therefore have an experimental {\it proof} of a high degree 
of coherence in multi-particle production in nuclear collisions at RHIC energies.  
The diagrams which allow to evade the AGK theorem are Gribov's ``inelastic
shadowing'' corrections \cite{glaubergribov} which correspond to the excitation of high--mass 
intermediate states in multiple scattering; the process thus no longer can be decomposed in terms 
of elementary $NN$ interactions. In parton language, these contributions correspond to 
multi--parton coherent interactions. 
\begin{figure}[htb]
\epsfig{file=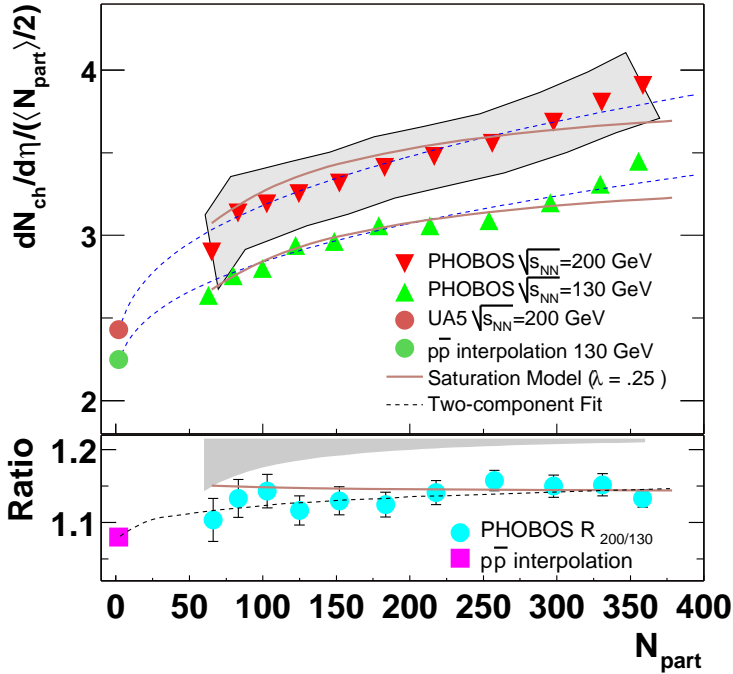,width=11cm}
\caption{Centrality dependence of the charged particle multiplicity near  mid-rapidity in Au + Au 
collisions at $\sqrt{s} = 130$ GeV and $200$ GeV, from \cite{Back:2002uc}.}
\label{fig:phobcolor}
\end{figure}

\subsubsection{Centrality dependence}

The dependence of multiplicity upon the number of participants discussed above can be established by 
selecting different centrality cuts. The result is shown in Fig.\ref{fig:phobcolor}; one can see that 
the multiplicity per participant pair increases with centrality, but not quite as fast as it would 
if the $NN$ collisions were independent. 

If we decompose the multiplicity measured in $NN$ collisions 
at some energy $\sqrt{s}$ into a fraction $X(s)$ coming from ``hard'' processes, 
and the remaining fraction $1-X(s)$ coming from 
``soft'' processes, and assume that in nuclear collisions ``hard'' processes are incoherent and thus 
scale with the number of collisions, whereas ``soft'' processes 
scale with the number of participants \cite{Bialas:1976ed}, we arrive at the following simple parameterization 
\cite{Kharzeev:2000ph}
\beq
{d n_{AA} \over d \eta} = \left[ (1-X(s))\ \left< N_{part} \right> + X(s) \left< N_{coll} \right>\right] \ 
{d n_{NN} \over d \eta},
\eeq
which describes the data quite well. In the framework of perturbative QCD approach, one has to assume 
that the coefficient $X(s)$ is proportional to the mini--jet production cross section, and thus 
grows with energy reflecting the growth of the parton distributions at small Bjorken $x$, 
$X(s) \sim [x G(x)]^2$, with $x \sim 1/\sqrt{s}$. Therefore one expects \cite{Wang:2000bf} that 
the centrality dependence should become increasingly steep as the $\sqrt{s}$ increases (for the latest 
development, see however \cite{Li:2001xa}). This increase 
is not seen in Fig.\ref{fig:phobcolor}, which in the lower panel shows that the ratio of the  
distributions at $\sqrt{s} = 200$ GeV and  $\sqrt{s} = 130$ GeV is constant within error bars.
The almost constant ratio appears consistent with the prediction \cite{Kharzeev:2001gp},\cite{Kharzeev:2000ph} 
based on the ideas of parton saturation, where the increase of multiplicity stems from the running of the 
QCD coupling constant determining the occupation number $\sim 1/\alpha_s$ of gluons in the classical field.   
The forthcoming results at $\sqrt{s} = 20$ GeV will further constrain the production dynamics. 

\subsubsection{Rapidity distributions}

\begin{figure}[htb]
\epsfig{file=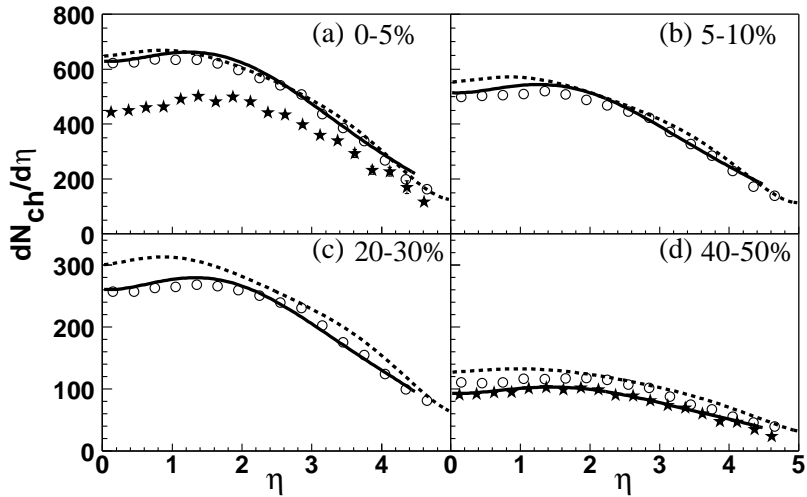,width=11cm}
\caption{Pseudo-rapidity distributions of charged particles from Au + Au  collisions at 
$\sqrt{s} = 200$ GeV (open circles), from \cite{Bearden:2001qq}. Solid line is the prediction based 
on parton saturation \cite{Kharzeev:2001gp}, and dashed line is the multi-phase transport 
model calculation \cite{Zhang:1999bd}. The $\bar{p}p$ distribution rescaled by $\langle N_{part}/2 \rangle$ is 
shown by stars.}
\label{fig:brahmsrap}
\end{figure}

Distributions of the produced particles in the emission angle $\theta$ (with respect to the collision axis), 
or pseudo-rapidity $\eta = - \ln [\tan (\theta/2)]$ provide another important characteristic of the 
collision process. Two features of RHIC results (see Fig. \ref{fig:brahmsrap}) are especially 
noteworthy: i) the distributions do not exhibit scaling in $\eta$; ii) the deviation from $NN$ results is 
maximal in the central rapidity region whereas the shapes of the $AA$ and $NN$ distributions are similar 
in the fragmentation region. 
Moreover, when corrected for the different beam rapidity $\eta \to \eta - y_{beam}$, distributions at 
different energies exhibit approximate scaling in the fragmentation region (``limiting fragmentation''). 

\subsubsection{Azimuthal distributions}

\begin{figure}[htb]
\epsfig{file=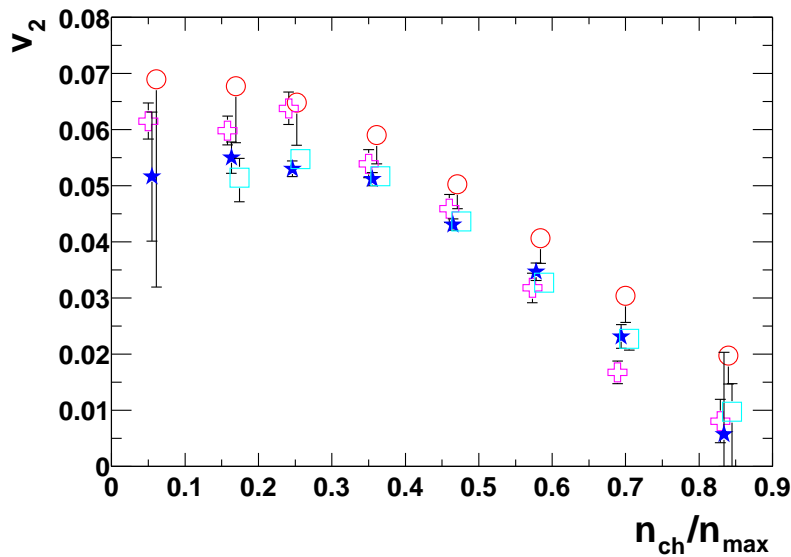,width=11cm}
\caption{Azimuthal anisotropy of hadron production in Au + Au collisions at at $\sqrt{s} = 130$ GeV; 
$v_2$ is the weight of the second harmonic, $\cos 2 \varphi$, in the particle distribution in the 
azimuthal angle $\varphi$; from \cite{Adler:2002pu}.}
\label{fig:v2star}
\end{figure}

Of great interest and importance are the distributions of the produced hadrons in the azimuthal angle. Indeed, 
if all of the $NN$ collisions were independent, there would be no reason to expect asymmetry in the 
distribution of the produced hadrons in the azimuthal angle (measured with respect to the reaction plane). 
This is why the azimuthal asymmetry represents a sensitive test of the collective effects in nuclear 
collisions. The azimuthal angle distributions of the produced hadrons are usually expanded in harmonics in the 
following way\footnote{The absence of the terms proportional to $\sin n \varphi$ is the consequence of parity conservation; 
it would be interesting to search for their presence in the data in view of the speculative scenarios allowing for 
$P$ and $CP$ violation in hot QCD \cite{Kharzeev:1998kz}.}:
\beq
{d N \over d \varphi} = 1 + 2 v_1 \cos \varphi + 2 v_2 \cos 2 \varphi + ...
\eeq
Fig. \ref{fig:v2star} shows the extracted from RHIC data coefficient $v_2$ ($v_2 \neq 0$ in the language of the 
field corresponds to ``elliptic flow''). One can see that the asymmetry of the azimuthal distribution is 
quite sizable, and for peripheral collisions (small multiplicity $n_{ch}/n_{max}$) reaches about $35 \%$. 

This effect certainly indicates the presence of collectivity in nuclear collisions, and comes out 
quite naturally in hydrodynamical calculations which assume complete thermalization in the final state \cite{Teaney:2000cw}, 
\cite{Kolb:2000fh}. However, final state effects are not the only possible origin of the azimuthal asymmetry; 
indeed, as we have discussed above, the high degree of coherence in the {\it initial} state signaled by the 
multiplicity measurements, in the parton saturation scenario, introduces strong correlation between the 
transverse 
momentum of the parton and its transverse coordinate in the wave functions of the nuclei. When the nuclei 
collide, this effect mimics at least a part of the asymmetry usually ascribed exclusively 
to final--state interactions 
\cite{Kovchegov:2002nf}, \cite{Krasnitz:2002ng}. The magnitude of the elliptic flow which can originate from 
the coherence in the initial state is still a subject of ongoing research and debates.      

%\begin{figure}[htb]
%\epsfig{file=v2starpt.eps,width=13cm}
%\caption{The transverse momentum dependence of the azimuthal anisotropy of hadron production in Au + Au collisions at at $\sqrt{s} = 130$ GeV; from \cite{Adler:2002pu}.}
%\label{fig:v2starpt}
%\end{figure}

The dependence of the elliptic flow on the transverse momentum of the hadrons presents a puzzle 
 \cite{Adler:2002pu}; 
the value of $v_2$ is seen to first increase with $p_t$, and then saturate. This 
contradicts to hydrodynamics, predicting a monotonic increase of $v_2$ with $p_t$; of course 
hydrodynamics cannot be trusted at high $p_t$ anyway because the density of hard particles is 
too small to allow for a meaningful statistical description. Energy loss of the produced 
jets can contribute to the azimuthal asymmetry at high $p_t$  (see, e.g., \cite{Gyulassy:2001kr}), 
even though it is not yet clear if the magnitude of the effect can be reproduced under realistic 
assumptions about the density of the medium and the jet interaction cross section  \cite{Shuryak:2001me}. 

\subsubsection{Hadron abundances}

The measurements of yields of different hadrons at RHIC hold many surprises. Of particular interest is  
the fact that even at RHIC energies the asymmetry between baryons and antibaryons 
is still sizable, with $\bar{p}/p$ ratio about $0.65$ \cite{Adler:2001bp}. This signals the diffusion of 
baryon number to quite small values of Bjorken $x \sim 10^{-2}$.  
 
\subsection{Hard processes}
\vskip0.3cm
\subsubsection{Suppression of high $p_t$ particles}
\begin{figure}[htb]
\epsfig{file=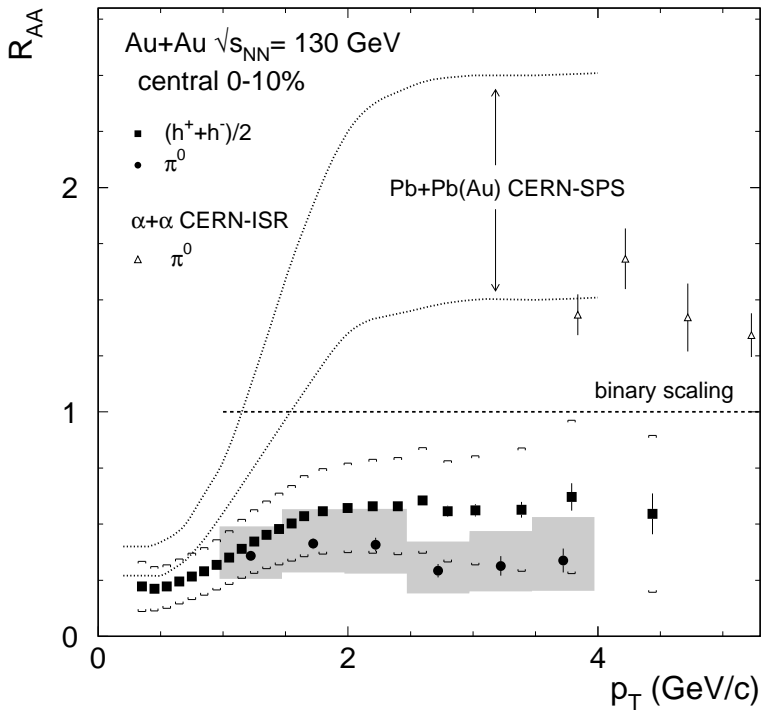,width=11cm}
\caption{The ratio of transverse momentum distributions of charged hadrons and neutral pions in 
Au + Au and $pp$ collisions at $\sqrt{s} = 130$ GeV; from \cite{Adcox:2001jp}.}
\label{fig:phenixpt}
\end{figure}
Jet energy loss was among the first signatures suggested for the 
diagnostics of the hot quark--gluon matter \cite{Bj,GW,BDMPS}. This is why the measurements 
of the high $p_t$ hadron production excited a lot of interest. Indeed, the experimental 
results are striking -- 
as can be seen in Fig.\ref{fig:phenixpt}, the yield of high $p_t$ hadrons is drastically reduced with respect 
to what is expected for incoherent production in $NN$ collisions. This behavior is very different from what 
was previously observed in $Pb-Pb$ collisions at CERN SPS and in $\alpha-\alpha$ collisions at CERN ISR 
(see  Fig.\ref{fig:phenixpt}). Does this important discovery signal jet energy loss in the quark--gluon plasma? 
The answer to this question can be given after we know the results of 
the forthcoming measurements in $p(d) A$ collisions, which would allow to separate clearly the effects 
coming from the initial state.  
 
%\begin{figure}[htb]
%\epsfig{file=phenixpt.eps,width=13cm}
%\caption{Transverse momentum distributions of charged hadrons (left) and neutral pions (right) in 
%Au + Au collisions at $\sqrt{s} = 130$ GeV; from  \cite{Adcox:2001jp}.}
%\label{fig:phenixpt}
%\end{figure}

\subsubsection{$B/\pi$ puzzle}

Another striking puzzle at RHIC is a rapid increase of the baryon--to--pion ratio in central 
$Au-Au$ collisions at high $p_t$ \cite{Adcox:2001mf}. The growth of this ratio is expected in the 
hydrodynamical scenario, in which equal velocity of the expanding parton ``fluid'' implies higher 
transverse momentum for more massive particles. However, the validity of hydrodynamical description 
is dubious at high $p_t$ where the density of particles is too small.  
If we assume that minijet fragmentation is the leading production mechanism of high $p_t$ particles, then 
the growth of the $B/\pi$ ratio implies that minijet fragmentation is severely affected by the medium. Another 
scenario \cite{Vitev:2001zn} attempts to explain both $B/\pi$ puzzle and a large value of baryon 
asymmetry $\bar{B}/B \neq 1$, and invokes the contribution of non--perturbative gluonic junctions in nuclear 
collisions \cite{Kharzeev:1996sq}. 

\subsubsection{Charm production}
\begin{figure}[htb]
\epsfig{file= 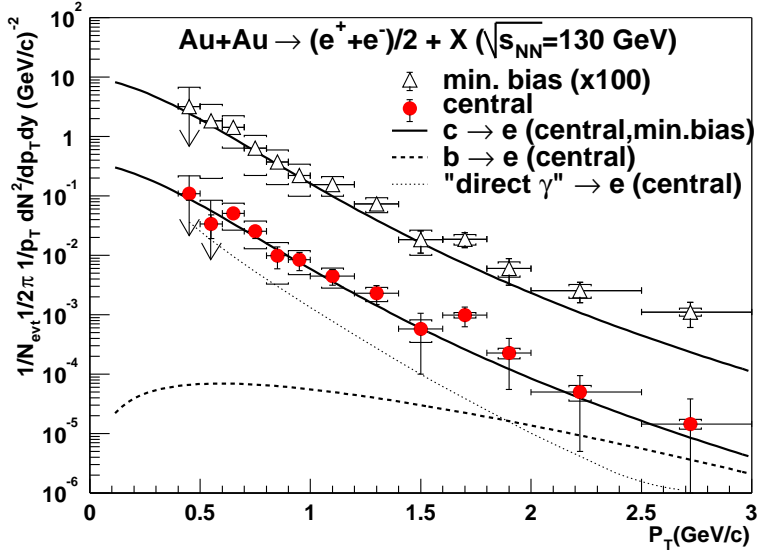,width=11cm}
\caption{The background--subtracted electron spectra for minimum bias ($0-92 \%$) and central ($0-10 \%$) 
collisions compared with the expected contributions from open charm decays; from \cite{Adcox:2002cg}.}
\label{fig:phencharm}
\end{figure}
The production of heavy flavors and quarkonia represents a very important and exciting part of RHIC program. 
While these studies will benefit in the future from increased luminosity and improvements in the detectors 
(allowing, in particular, to reconstruct the decay vertex of heavy hadrons), the first measurement of 
charm production cross section has been already reported \cite{Adcox:2002cg}. This has been done by deciphering the 
charm decay contribution to the single electron spectrum -- see Fig.\ref{fig:phencharm}. 
Of particular interest is the fact that while the production cross sections of light hadrons, as discussed above, 
show strong nuclear effects, the cross section of charm production, within the error bars of the measurement, scales 
with the number of $NN$ collisions. 
These results may imply much smaller, in comparison to light quarks, energy loss of heavy quarks \cite{Dokshitzer:2001zm} 
stemming from the suppression of the gluon radiation at small angles (``dead cone effect'').

\section{What have we learned so far?}

It is too early to assess the implications of RHIC results; however, it is becoming increasingly clear that 
they challenge most, if not all, of the existing theoretical dogmas. A coherent and convincing theory 
describing all of the observed phenomena directly in terms of QCD still has to be born. 
However, we can already conclude that many of the observed phenomena clearly manifest collective behavior; 
nuclear collisions at RHIC are not an incoherent superposition of nucleon--nucleon collisions.
 
The measured particle multiplicities and transverse momentum spectra allow to estimate initial 
energy density at the early moments of the collision; a typical value inferred in this way is about 
$20\ {\rm GeV/fm}^3$ (see, e.g., \cite{Kharzeev:2000ph}). The dynamics of strongly interacting matter at 
such energy density (exceeding the energy density in a nucleus by over two orders of magnitude!) 
should be described in terms of quarks and gluons, and the collective phenomena observed at RHIC thus 
directly reflect the properties of high density QCD.

\end{document}